\documentclass{aa2}
\usepackage{graphicx}
\usepackage{times}
\newcommand{\etal}{et al.~}
\newcommand{\msun}{{\,\rm M}_{\odot}}

\newcommand{\nht}{\ifmmode {{\rm NH}_3} \else {NH{\bas 3}} \fi}
\newcommand{\tco}{\ifmmode {^{13}{\rm CO}} \else {$^{13}{\rm CO}$}\fi}
\newcommand{\dco}{\ifmmode {^{12}{\rm CO}} \else {$^{12}{\rm CO}$}\fi}
\newcommand{\cdo}{\ifmmode {{\rm C}^{18}{\rm O}} \else {${\rm C}^{18}{\rm O}$}\fi}

\newcommand{\htco}{\ifmmode {{\rm H}^{13}{\rm CO}^{+}} \else {${\rm H}^{13}{\rm CO}^{+}$}\fi}
\newcommand{\hco}{\ifmmode {{\rm H}^{12}{\rm CO}^{+}} \else {${\rm H}^{12}{\rm CO}^{+}$ }\fi}

\newcommand{\jtd}{\ifmmode {{\rm J}=3\!\rightarrow\!2} \else {${\rm J}=3\!\rightarrow\!2$} \fi}
\newcommand{\jcq}{\ifmmode {{\rm J}=5\!\rightarrow\!4} \else {${\rm J}=5\!\rightarrow\!4$} \fi}
\newcommand{\as}{\ifmmode {^{\scriptscriptstyle\prime\prime}}
\else $^{\scriptscriptstyle\prime\prime}$\fi}
\newcommand{\am}{\ifmmode {^{\scriptscriptstyle\prime}}
\else $^{\scriptscriptstyle\prime}$\fi}
\begin{document}
\title{The Observation of Circumstellar Disks: Dust and Gas Components}
\author{Anne Dutrey \inst{1}, Alain Lecavelier des Etangs \inst{2} \& Jean-Charles Augereau \inst{3,2}}
\institute{$^1$ Observatoire de Bordeaux, \\ 2 rue de l'Observatoire, F-33270 Floirac, France \\
$^2$ Institut d'Astrophysique de Paris (IAP) \\ 98bis boulevard Arago, F-75014 Paris, France \\
$^3$ Service D'Astrophysique du CEA-Saclay \\ F-91191
Gif-sur-Yvette cedex, France}
\date{}
\authorrunning{}
\titlerunning{}
\abstract{Since the 1990's, protoplanetary disks and planetary disks
have been intensively observed from the optical to the millimetre
wavelength and many models have been developed to investigate their
gas and dust properties and dynamics. These studies remain empirical
and rely on poor statistics with only a few well known
objects. However, the late phases of the stellar formation are among
the most critical for the formation of planetary systems. Therefore,
we believe it is timely to tentatively summarize {\it the observed
properties of circumstellar disks around young stars from the
protoplanetary to the planetary phases}. Our main concern is to
present the physical properties considered as observationally robust
and to show their main physical differences associated to an
evolutionary scheme. We also describe areas still poorly understood
such as how protoplanetary disks disappear to lead to planetary disks
and eventually planets.}

\maketitle{}
\section{Introduction}
Before the 1980's, the existence of protoplanetary disks of gas
and dust around stars similar to the young Sun (4.5 billion years
ago) was inferred from the theory of stellar formation (e.g.
Shakura \& Suynaev 1973), the knowledge of our own planetary
system and dedicated models of the Proto-Solar Nebula. The
discovery of the first bipolar outflow in L1551 in 1980
drastically changed the view of the stellar formation. In the
meantime, optical polarimetric observations by Elsasser \& Staude
(1978) revealed the existence of elongated and flattened
circumstellar dust material around some Pre-Main-Sequence (PMS)
stars such as the low-mass TTauri stars. The TTauri are understood
to be analogs to the Sun when it was about $10^6$ years old. A few
years later, observations from the InfraRed Astronomical Satellite
(IRAS) found significant infrared (IR) excesses around many TTauri
stars, showing the existence of cold circumstellar dust (Rucinski
1985). More surprisingly, IRAS also show the existence of weak IR
excess around Main-Sequence (MS) stars such as Vega, $\epsilon$
Eridani or $\beta$ Pictoris (Aumann \etal 1984). These exciting
discoveries motivated several groups to model the Spectral Energy
Distribution (SED) of TTauri stars (e.g. Adams \etal 1987) and
Vega-like stars (Harper \etal 1984).

On one hand, for PMS stars, the emerging scenario was the
confirmation of the existence circumstellar disks orbiting the
TTauri stars, the gas and dust being residual from the molecular
cloud which formed the central star (Shu \etal 1987). Since such
disks contain enough gas (H$_2$) to allow, in theory, formation of
giant planets, they are often called ``protoplanetary'' disks.
During this phase, the dust emission is optically thick in the
Near-IR (NIR) and the central young star still accretes from its
disk. Such disks are also naturally called accretion disks.

On the other hand, images of $\beta$ Pictoris by Smith \& Terrile
(1984) demonstrated that Vega-type stars or old PMS stars can also
be surrounded by optically thin dusty disks. These disks were
called ``debris'' disks or later ``planetary'' disks because
planetesimals should be present and indirect evidence of planets
was found in some of them ($\beta$~Pic).

In this chapter, we review the current observational knowledge of
circumstellar disks from the domain of the UV to the millimeter
(mm). We discuss in Sections \ref{ttauri} and \ref{herbigae} the
properties of protoplanetary disks found around young low-mass
(TTauri) and intermediate-mass (Herbig AeBe) stars. In section
\ref{transition}, we summarize the properties of {\it transition
disks} which have still some gas component but also have almost
optically thin dust emission in the NIR, objects which are thought
to be in the phase of dissipating their primary gas and dust. We
present the properties of optically thin dust disks orbiting old
PMS, Zero-Age-Main-Sequence (ZAMS) or Vega-type stars, such as the
$\beta$ Pic {\it debris disk} in Section \ref{planeto}. We
conclude by reviewing future instruments and their interest for
studying such objects.
\label{intro}
%
\section{Protoplanetary disks: The TTauri stars}
\label{ttauri}
Following the standard classification (see Chapter 4.2.4), TTauri
 stars typically present the SEDs of Class II objects.

Evidence for disk features around these young stars comes
principally from the following observational considerations:

\begin{enumerate}

\item A flat and geometrically thin distribution account for the
SED (produced by the dust emission) from the optical to the mm
(including the IR excess) because the extinction towards most
TTauri stars is very low.

\item In the 1990's, adaptive optics (AO) systems on ground-based
telescopes and the Hubble Space Telescope (HST) started to image
these disks. Dust grains at the disk surface scatter the stellar
light, revealing the disk geometry of circumstellar material as in
the case of HH30 (Fig.\ref{hh30}).

\item Large millimeter arrays such as OVRO or the IRAM
interferometers clearly demonstrate by mapping the CO J=1-0 and
J=2-1 line emission from the gas that the circumstellar material
has a flattened structure and is in Keplerian rotation.

\end{enumerate}

Most TTauri stars form in binary or multiple systems (Mathieu
\etal 2000) and many observational results show that binarity
strongly affects the dust and gas distribution as a result of
tidal truncations. The material can be in a circumbinary ring as
in the GG Tau disk (Dutrey \etal 1994) or confined in small,
truncated, circumstellar disks. However, for simplicity, we will
focus here on properties of disks encountered around stars known
as single.

\subsection{Mass Accretion rates}

TTauri stars with IR excess usually present optical emission lines
(e.g. Edwards \etal 1994). Studies of these lines reveal that the
stars are still accreting/ejecting material from their disk even
if the main ejection/accretion phase is over (see Chapter 4). When
they present strong H$_\alpha$ emission lines (equivalent
linewidth $W_{H_\alpha} \geq 10$ \AA), they are called Classical
Line TTauri stars (CTTs).

Observations show that the mass accretion rate (and also the mass
ejection rate) decreases from the protostars to the Class III
phases by several order of magnitudes (Hartmann \etal 1998,
Chapter 4). Despite the uncertainties resulting from the various
observational methods and tracers used for measuring both rates,
there is a clear correlation between mass loss and mass accretion.
Typical values for the mass accretion rate of a few $10^{-5}
\msun$/yr are found for Class 0 objects while TTauri stars have
lower values around $10^{-8} \msun$/yr. Some protostars such as
the FU Orionis objects even exhibit episodic outbursts with
accretion rates as high as a few $10^{-4} \msun$/yr (Hartman \&
Kenyon 1996). Assuming a mass accretion rate of $ \sim
10^{-8}\msun$/yr, a TTauri star of 0.5~$\msun$ would accrete only
0.01~$\msun$ in 1~Myr. Hence, most of the accretion must occur in
the protostellar phase.

\subsection{Modelling the Spectral Energy Distribution}

Global properties of disks can be inferred from the SED but, due
to the lack of angular resolution, the results are strongly model
dependent and there is usually no unique solution. Moreover many
stars are binaries and SEDs are not always individually resolved,
leading to possible misinterpretations.

 Small dust particles (of radius $a$) are
efficient absorbers of wavelengths radiation with $\lambda \leq
a$. In equilibrium between heating and cooling they re-emit at
longer wavelengths a continuous spectrum which closely resembles a
thermal spectrum. At short wavelengths, the scattering of the
stellar light by dust grains can dominate the spectrum, the limit
between the scattering and the thermal regimes being around $\sim
3-5 \mu$m. Very close to the star, ($\sim 0.1 - 5$~AUs) the
temperature is high enough ($\sim 500-1000$~K) and the NIR/optical
continuum can be dominated by the thermal emission of very hot
grains.

Spectral Energy Distributions can be reproduced by models of disks
(e.g., Pringle 1981, Hartman 1998) which assume that i) the disk
is reprocessing the stellar light (passive disk) or ii) the disk
is heated by viscous dissipation (active disk). In both models,
since there is no vertical flow, the motions are circular and
remain Keplerian ($v(r) = \sqrt{GM_*/r}$, where $M_*$ is the
stellar mass). As a consequence the disk, in hydrostatic
equilibrium, is geometrically thin with $H << r$ where $H$ is the
disk scale height. For viscous disks, the viscosity $\nu$ is
usually expressed by the so-called $\alpha$ parameter linked to
$\nu$ by $\nu = \alpha c_s H$ where $c_s$ is the sound speed. The
accretion remains subsonic with $\nu/r \sim \alpha c_s H/r << c_s$
and $\alpha \sim 0.01$.

Chiang \& Goldreich (1997) have also developed a model of passive
disk where the optically thin upper layer of the disk is
super-heated above the blackbody equilibrium temperature by the
stellar light impinging the disk which produces a kind of disk
atmosphere. Both viscous heating and super-heated layers seem to
be necessary to properly take into account the observations
(D`Alessio \etal 1998, 1999).

Many observers interpret the SEDs assuming that the disk is
geometrically thin with simple power law dependencies versus
radius for the surface density ($\Sigma(r) = \Sigma_o
(r/r_o)^{-p}$) and the temperature ($T_k(r) = T_o (r/r_o)^{-q}$).
In this case, there is no assumption about the origin of the
heating mechanism, and the results are compared to more
sophisticated models similar to those described above.

Dust disks are usually optically thick up to $\lambda \sim 100-200
\mu$m, allowing us to trace the disk temperature. In active disks
as in passive geometrically thin disks, the radial dependence of
the temperature follows $T_k(r) \propto r^{-0.75}$. Beckwith \etal
(1990) have found that typical temperature laws encountered in
TTauri disks are more likely given by $\propto r^{- 0.65- 0.5}$.
However Kenyon \& Hartman (1987) have shown that for a flaring
disk, the temperature profile should be as shallow as $q=0.5$,
closer to the observed values.

\subsection{The dust content}

Longward of $\sim 100 \mu$m wavelength, the dust emission becomes
optically thin. Observations, at these wavelengths, are well
explained by a dust absorption coefficient following $\kappa_\nu =
\kappa_o (\nu/10^{12}\rm[Hz])^\beta$ with $\beta \simeq 0.5-1$ and
$\kappa_o = 0.1 $ cm$^{2}$g$^{-1}$ of gas + dust (with a gas to
dust ratio of 100, Beckwith \etal 1990). The spectral index is
significantly lower than in molecular clouds where $\beta \simeq
2$. Compared to values found in molecular clouds, both $\beta$ and
$\kappa_o$ show that a significant fraction of the grains have
evolved and started to aggregate (Henning \& Stognienko 1996,
Beckwith \etal 2000), grains may even have fractal structures
(Wright 1987). Dust encountered in protoplanetary disks seems to
be a mixture of silicate and amorphous carbon covered by icy
mantles (Pollack \etal 1994). The exact composition is poorly
known, recent VLT observations of broad-band absorption features
in the NIR start to put quantitative constrains on solid species
located on grain mantles such as CO or H$_2$O (Dartois \etal 2002,
Thi \etal 2002). These results also confirm that many molecules
may have condensed from gas phase on grains (see also Section 2.4
and 3) in the cold ($\sim 20-15$~K) outer part of the disk. Very
close to the star, the dust mantles may be significantly different
since the disk is hotter ($\sim 1000$~K at 0.1~AU).

Optical/NIR interferometry is a powerful tool to trace the very
inner disk. Monnier \& Millan-Gabet (2002) have observed several
disks around TTauri and Herbig Ae/Be stars. They found that the
observed inner disk sizes ($r_{in} \sim 0.1$~AU) of TTauri stars
are consistent with the presence of an optically thin cavity for a
NIR emission arising from silicate grains of sizes $a \geq
0.5-1$~$\mu$m and which are heated close to their temperature of
sublimation.

At NIR and optical wavelengths, small dust particles also scatter
the stellar light impinging the disk surface, producing reflection
nebulas imaged by optical telescopes. Fig \ref{hh30} shows, in
false color, the HH30 disk seen edge-on which appears as a dark
lane. The star ejects a jet perpendicular to the disk plane and is
highly obscured by the material along the line-of-sight (visual
extinction up to $A_v \geq 30$ mag). Only the disk surface or
atmosphere (Chiang \& Goldreich 1997) is seen. Due to the high
opacity, these data cannot allow us to estimate the dust mass
distribution without making {\it a priori} (or external)
assumptions about the vertical distribution. However, grains of
typical size around $a \sim 0.05 -1 \mu$m are responsible for
scattering (Close \etal 1998, Mac Cabe \etal 2002). Since forward
scattering is easier to produce than the backward scattering
(e.g., GG Tau, Roddier \etal 1996), the disk inclination usually
provides simple explanation for the observed brightness asymetry.

Estimating the gas and dust mass of these disks is done by several
methods (see also Section 2.4) but is quite uncertain. Analyzing
the SEDs in the optically thin part of the spectrum leads to
TTauri disk masses (gas+dust) ranging from 0.1 to 0.001 $\msun$
(Beckwith \etal 1990). These determinations suffer from many
uncertainties such as the value of $\kappa_o$ and the gas-to-dust
ratio which is usually assumed to have its interstellar value of
100. Moreover, the inner part of the disk is still optically thick
(up to radii of $\sim 10-30$~AUs at 3mm). Resolved images of the
thermal dust emission obtained with mm arrays allow a separation
of possible opacity and spectral index effects. This procedure
also estimates the surface density radial profile $\Sigma(r) =
\Sigma_0 (r/r_0)^{-p}$. Typical values of $p$ are around $1-1.5$
(Dutrey \etal 1996). Such low values imply that {\it the reservoir
of the mass is in the outer disk traced by sub-mm images}.
Assuming a single surface density distribution from $0.1$ to
$500$~AU, a disk with $p=1$ has only 10 \% of its mass located
within $r=10$~AU while with $p=1.9$, the same disk has 50 \% of
its mass within the same radius.

\vspace{0.3cm}

In summary, a significant fraction of the grains in protoplanetary
disks appears to be more evolved than in molecular clouds and
coagulation processes have already started. The vertical dust
distribution is however not yet constrained. Moreover,
observations at a given frequency are mainly sensitive to grains
of size $a \leq$ a few $\lambda$ (absorption or diffusion
cross-sections cannot significantly exceed the geometrical
cross-section, even for more complex grain features and
aggregates, Pollack \etal 1994, Kr\"ugel \& Siebermorgen 1994).
Only multi-wavelength analysis of resolved images, from the
optical to the cm domains, should allow to conclude about the dust
sedimentation. Estimating the mass of the disks remains difficult.
However, continuum mm images suggest that the reservoir of mass is
located at large distance $r \geq 30-50$~AUs. Using a gas-to-dust
ratio of 100, analyzes of the dust emission indicate that the
total (dust+gas) disk masses are in the range 0.001 to 0.1
$\msun$.

\subsection{The Gas Content}

In protoplanetary disks, molecular abundances are defined with
respect to H$_2$ since this is the main (gas) component and the
gas-to-dust ratio, which is not yet measured, is assumed to be
100, as in the molecular clouds. Several groups (Thi \etal 2001,
Richter et al. 2002) have recently started direct investigation of
H$_2$ (which only possesses quadrupolar moment) but the most-used
tracer of the gas phase remains CO.

Carbon monoxide is the most abundant molecule after molecular
hydrogen. Its first rotation lines are observable with current mm
interferometers allowing astronomers to trace the properties of
outer gas disks. Current sensitivity of mm arrays is limited and
does not allow the observation of CO lines for $r \leq 30-50$~AUs.
Since the density is very high ($n(\mathrm{H}_2) \geq 10^6$
cm$^{-3}$), the J=1-0 and J=2-1 CO lines are thermalized by
collision with H$_2$ in the whole disk. Hence, a simple model of
Keplerian disk assuming LTE conditions is sufficient to derive the
CO disk properties (Dutrey \etal 1994).

CO maps reveal that disks are in Keplerian rotation (Koerner \etal
1993) and that many disks in Taurus-Auriga clouds are large with
typical radii $R_{out} \simeq 300-800$~AUs (see Simon \etal 2000,
their Fig.1). Comparing resolved CO maps to disk models by
performing a $\chi^2$ minimization of the disk parameters
(Guilloteau \& Dutrey 1998, see their Fig.1) provides very useful
information about the density and temperature distributions. The
temperature radial profiles deduced from $^{12}$CO images are
consistent with stellar heating in flared disks and the turbulence
appears to be small, less than $0.1$~km.s$^{-1}$ (Dutrey \etal
2004). Since the $^{12}$CO and $^{13}$CO J=1-0 and J=2-1 have
different opacities, they sample different disk layers. A global
analysis of these lines permits one to derive the vertical
temperature gradient. Dartois \etal 2003 have shown that in the DM
Tau disk, the ``CO disk surface'', traced by $^{12}$CO, is located
around $\sim$ 3$H$ above the disk mid-plane, the $^{13}$CO J=2-1
samples material at about 1$H$ while J=1-0 is representative of
the disk mid-plane. They also deduce a vertical kinetic gradient
which is in agreement with disk models (e.g. D'Alessio \etal
1999), the mid-plane is cooler ($\sim 13$~K) than the CO disk
surface ($\sim 30$~K at 100~AU). This appears in the region of the
disk where the dust is still optically thick to the stellar
radiation while it is already optically thin to its own emission,
around $r \sim 50 - 200$~AU in the DM Tau case. Beyond $r \geq
200$~AU where the dust becomes optically thin to both processes,
the temperature profile appears isothermal vertically.

A significant fraction of the DM Tau disk has a temperature below
the CO freeze out point (17~K) but there remains enough CO in the
gas phase to allow the J=2-1 line of the main isotope to be
optically thick. The chemical behavior of molecules and coupling
between gas and dust are poorly known. There have only been a few
attempts to survey many molecules in protoplanetary disks (Dutrey
\etal 1997, Kastner \etal 1997, Zadelhoff \etal 2001). Today, in
addition to $^{13}$CO and C$^{18}$O, only the more abundant
species after the carbon monoxyde are detectable, like HCO$^+$,
CS, HCN, CN, HNC, H$_2$CO, C$_2$H and DCO$^+$. By studying the
excitation conditions of the various transitions observed in the
DM Tau disk, Dutrey \etal (1997) deduced molecular abundances
indicating large depletion factors, ranging from 5 for CO to 100
for H$_2$CO and HCN, with respect to the abundances in the TMC1
cloud. They also {\it directly} measured the H$_2$ density; the
total disk mass they estimated is a factor 7 smaller than the
total mass measured from the thermal dust emission. Both results
suffer from several uncertainties, only a more detailed analysis
will allow one to conclude that the gas-to-dust ratio is lower
than 100, even if such a behavior is expected.

Due to limited sensitivity, the chemistry of the gas phase in the
inner disk is poorly constrained (Najita \etal 2000). Models of
nebulae irradiated by stellar radiation, including X ray emission,
suggest a complex chemistry (Glassgold \etal 1997), even at
relatively large radii ($r \geq 50$~AUs, Najita \etal 2001).

\vspace{0.3cm}

\begin{figure}
\resizebox{8.0cm}{!}{\includegraphics[angle=270]{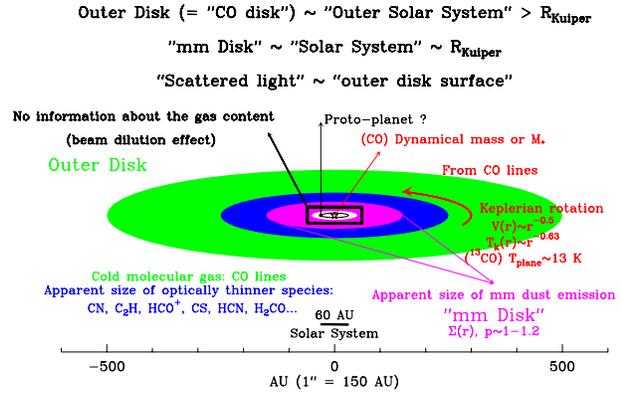}}
\caption[]{Observable properties of a protoplanetary disk
surrounding a TTauri star, located at D= 150~pc. This montage
presents the information currently derivable from observations,
and gives the ``apparent'' sizes for optically thick and thin
lines and thermal dust emission observed at mm wavelengths with
interferometers.} \label{protodisk}
\end{figure}

Fig.\ref{protodisk} summarizes the observable properties of a
protoplanetary disk encountered around a TTauri star of $0.5
\msun$ and located at a distance of 150~pc.

\subsection{Illustration through HH\,30}

The HH\,30 observations, shown Fig.\ref{hh30}, give one of the
most complete pictures of the material surrounding a PMS star.
Optical and mm observations are clearly tracing the same physical
object. In Fig.\ref{hh30}, the HH30 dust disk observed in the
optical by the HST (Burrows \etal 1996) is close to edge-on and
appears as a dark lane, only the disk surface or atmosphere is
bright. The jet emission is also seen, perpendicular to the disk
plane. Pety \etal (2003) have superimposed in contours to this
image the blueshifted and redshifted integrated emission (with
respect to the systemic velocity) of the $^{13}$CO J=2-1 line in
the disk. The CO disk extends as far as the dust disk and the
velocity gradient is along the major disk axis, as expected for
rotation. $^{12}$CO J=2-1 emission is also observed within the jet
(extreme velocity). In the HH\,30 case the low angular resolution
of the data does not allow one to separate between the $^{12}$CO
J=2-1 emission associated to the outflow, the cloud and
 the disk. This is a common problem which cannot be fully solved
  by current interferometric observations. Selecting sources which
are located in a region devoid of CO emission of the molecular
cloud minimizes the confusion.

 Since the disk is seen edge-on, the
vertical distribution can be estimated from the optical
observations of the dust (Burrows \etal 1996). The results are
somewhat model dependent but one can conclude that the disk is
pressure supported (dominated by the central star) with the best
fit given by $H(r) \propto r^{1.45}$. The authors estimate the
surface density law to be $\Sigma(r) \propto 1/r$ and the disk
mass $\sim 6.10^{-3} \msun$. CO observations reveal that the disk
is in Keplerian rotation around a central star of mass 0.5~$\msun$
and has an outer radius of R$_{out} = 440$~AUs. Interestingly,
this is in agreement with the best fit of the optical data which
gives R$_{out} \sim 400$~AU.

\begin{figure}
\resizebox{8.0cm}{!}{
}
\caption[]{(Background in false color) A montage of the dust disk
emission from the HST (from Burrows \etal 1996) and its
perpendicular jet. Contours present $^{12}$CO J=2-1 emission
associated to the jet (black and white) corresponding to the
extreme velocity and the redshifted and blueshifted integrated
emission with respect to the systemic velocity of $^{13}$CO J=2-1
line coming from the disk (from Pety \etal 2003). Note that the
velocity gradient of the $^{13}$CO J=2-1 emission is along the
major disk axis, as expected for rotation. Both mm and optical
images are tracing two different aspects of the same physical
object and contribute to give a coherent outstanding of its
physical properties.} \label{hh30}
\end{figure}

\subsection{Proplyds}

The disks we described have so far are found in low-mass star
forming regions such as the Taurus-Auriga clouds. Physical
properties of disks surrounding low mass-stars born inside
clusters forming massive stars may be significantly different
because they are exposed to the strong ambient UV field generated
by nearby OB stars and can be photo-evaporated by it (Johnstone
1998). This is the case for the proplyds which are protoplanetary
disks seen in silhouette against the strong HII region associated
to the Trapezium Cluster in Orion~A ({\it e.g.} McCaughrean \&
O'Dell 1996).
%
\section{Disks around intermediate-mass stars: The Herbig Ae/Be stars}
\label{herbigae}
%
With masses in the range 2-8 $\msun$, Herbig Ae/Be (HAeBe) stars,
massive counterparts of TTauri stars, are the progenitors of A and
B Main Sequence stars. Since they are more luminous and massive
than the TTauri stars, the surrounding material is submitted to
stronger UV and optical stellar flux.

Several Herbig Ae stars are isolated but located in nearby star
forming regions (Taurus, R Oph); their observed properties can be
directly compared with those of TTauri stars. This is not the case
for Herbig Be stars because most of them are located at larger and
uncertain distance ($D \geq 500-800$~pc). Therefore, we will
discuss in this section mainly isolated Herbig Ae stars.

Like TTauri stars, SEDs of HAeBe stars exhibit strong IR excesses.
Optical and NIR observations (Grady \etal 1999) reveal that many
of these objects are surrounded by large reflection nebulae (e.g.
more than 1000 AUs large for the A0 star AB Auriga) revealing
envelopes or halos (Leinert \etal 2001). There is however now
clear evidences that Herbig Ae stars are also surrounded by disks.
In particular, resolved CO maps from mm arrays reveal that the
circumstellar material is also in Keplerian rotation (e.g. MWC480,
an A4 star: Manning \etal 1997 and HD\,34282, an A0 star: Pietu
\etal 2003). Millimetre continuum surveys also suggest that the
total surrounding mass may have a tendency to increase with the
stellar mass (see Natta \etal 2000, Fig.1). So far, one of the
more massive Keplerian disk ($\sim 0.11 \msun$) has been found
around an A0 Herbig Ae star: HD\,34282 (Pietu \etal 2003).

Since the medium is hotter than for TTauri stars, one would expect
different behavior and in particular a rich chemistry. The limited
sensitivity at present of molecular surveys at mm wavelengths does
not allow one to distinguish significant differences and outer
disks ($r \geq 50$~AUs) of Herbig Ae stars appears similar to
``cold'' outer disks found around TTauri stars.

However, most of the differences should appear in the warm
material, closer to the star. Optical/NIR interferometric
observations by Monnier \& Millan-Gabet (2002) revealed that the
observed inner radius of disks is usually larger for HAeBe stars
than for TTauri stars. This is understood in term of truncation by
 dust sublimation close to the star. They also found that grain
sizes are about similar for TTauri and HAeBe stars. Dullemond
\etal (2001) have shown that direct irradiation of Herbig Ae disks
at their inner radius can explain the bump observed at IR
wavelengths in the SEDs of Herbig Ae stars. In a few cases, as for
HD\,100546, direct detection of H$_2$ with FUSE (Lecavelier \etal
2003) reveals the existence of warm ($T \simeq 500$~K) molecular
gas close to the star ($r \sim 0.5- 1$~AUs).

The fact that HAeBe stars are hotter has also favored the use of
ISO to characterize the geometry and the dust composition of the
disk close to the star. Bouwman \etal (2000) have performed a
detailed spectroscopic study from 2 to 200~$\mu$m of the
circumstellar material surrounding AB Auriga (A0) and HD\,163296 (
A1). Their analysis of the SEDs, assuming an optically thin dust
model, has revealed the existence of both hot ($T\sim 1000$~K) and
cold ($T\sim 100$~K, most of the mass) dust components while the
NIR emission at 2$\mu$m can be explained the presence of metallic
iron grains. As in TTauri disks, substantial grain growth has
occured, with grain size up to $\sim$~0.1-1mm. It is also
important to note that comparisons of the ISO spectrum of the
Herbig Ae star HD\,100546 with those of the comet Hale-Bopp have
revealed many similarities (Waelkens \etal 1999).

\subsection{MWC\,480: similarities and differences with a TTauri disk}

MWC\,480 is located at $D = 140$~pc, in Auriga cloud. CO
observations by Manning \etal (1997) have revealed that the
surrounding disk is in Keplerian rotation around an A4 star (Simon
\etal 2000). The disk is large ($R_{out}\simeq 600$~AUs) and
inclined by about 35$^o$ along the line of sight. The stellar mass
is around $\sim 2 \msun$. The temperature deduced from the
optically thick $^{12}$CO J=2-1 line (which probes about 3 scale
heights above the disk mid-plane) is $T\sim 60$~K at $R =
100$~AUs, this is significantly larger than the temperature of
$\sim 30$~K found for TTauri disks using the same tracer.
Interestingly, the disk is not detected in the NIR in scattered
light. From the non-detection, Augereau \etal (2001a) have deduced
that either the dust emission at 1.6~$\mu$m is too optically thin
to be detected or there is a blob of optically thick material
close to the star which hides the outer disk to the stellar
radiation. Knowledge of the disk scale height is required to
decide between these possibilities. The existence of such blobs is
also favored by SED models of several Herbig Ae stars (Meeus \etal
2001, their Fig.8).

\vspace{0.3cm} NIR and CO/mm detections are not necessarily
linked: the TTauri DM Tau has the best known disk at mm
wavelengths but the disk was only recently detected in the NIR by
performing deep integration with the HST (Grady \etal 2003).
Keeping this in mind, the MWC480 disk appears very similar to a
TTauri disk, at least for the outer part ($r \geq 50$~AUs). It is
just hotter and maybe somewhat more massive.

\subsection{ The UX Ori phenomenon}

Some HAeBe stars, such as UX Orionis, have a very complex
spectroscopic, photometric and polarimetric variability which has
been, in some cases, monitored for years. The variability has
usually a short periodicity of order $\sim 1$ year and can be as
deep as two magnitudes in the V band.

It is very tempting to link this phenomenon to planetary
formation; the variability could be caused by clumps of material
(such as clouds of ``proto comets''), located in the very inner
disk and orbiting the star. Natta \& Whitney (2000) have developed
a model in which a screen of dust sporadically obscures the star,
this happens when the disk is tilted by about $45^o-68^o$ along
the line of sight. One clearly needs more sensitive
multi-wavelength data at high angular resolution on a large sample
of Herbig Ae stars to distinguish among the various models.
\section{From Protoplanetary to Planetary Disks}
\label{transition}
On one side, one finds massive {\it gaseous} protoplanetary disks
around TTauri and Herbig Ae stars and on the other, one finds {\it
dusty} planetary disks around young MS stars. A natural question is
then: {\it are there disks in an intermediate state and if so, what
are their observational characteristics?}

Limited in sensitivity by current telescopes, we know only a few
examples of objects which can be considered as ``transition'' disks.

\subsection{The surprising case of BP Tau}

BP Tau is often considered as the prototype of CTTs. It has a high
accretion rate of $\sim 3 \cdot 10^{-8} \msun$/yr from its
circumstellar disk which produces its strong excess emission in the
ultraviolet, visible and NIR (Gullbring \etal 1998). It is also very
young ($6 \cdot 10^{5}$yr, Gullbring \etal 1998).  Despite these
strong CTTs characteristics, its mm properties are very different than
those of other TTauri stars surrounded by CO disks.

Recent CO J=2-1 and continuum at 1.3mm images from the IRAM array
have revealed a weak and small CO and dust disk (Simon \etal
2000). With a radius of about $\simeq 120$~AUs, the disk is small
and is in Keplerian rotation around a $(1.3 \pm 0.2)
(D/140\mathrm{pc}) \msun$ mass star. A deeper analysis of these CO
J=2-1 data (Dutrey \etal 2003) also shows that the J=2-1
transition is marginally optically thin, contrary to what is
observed in other TTauri disks. The disk mass, estimated from the
mm continuum emission by assuming a gas-to-dust ratio of 100, is
very small $1.2 \cdot 10^{-3} \msun$, a factor 10 below the
minimum initial mass of the Solar Nebula, for comparison. By
reference to the mass deduced from the continuum, the CO depletion
factor can be estimated; this leads to a factor as high as $\sim
150$ with respect to H$_2$. Even taking into account possible
uncertainties such as a lower gas-to-dust ratio or a higher value
for the dust absorption coefficient, the CO depletion remains high
compared to other CO disks. Finally, the kinetic temperature
derived from the CO data is also relatively high, about $\sim
50$~K at 100 AUs.

Both the relatively high temperature and the low disk mass suggest
that a significant fraction of the disk might be superheated
(above the black body temperature) similarly to a disk atmosphere
(e.g. Chiang \& Goldreich 1997, see also Section 2). With
reasonable assumptions for the dust grain properties and surface
density, one can then estimate the fraction of small grains ($a
\simeq 0.1 \mu$m) still present in the disk to reach in the
visible $\tau_V =1$ at the disk mid-plane. Since it corresponds to
a total mass of small grains of about 10 \% of the total mass of
dust ($1.2\cdot 10^{-5} \msun$) derived from the mm continuum
data, this is not incompatible with the current data but should be
confirmed by optical and NIR observations.

It is also interesting that the CO content of BP Tau is too high
to result from evaporation of proto-comets (Falling Evaporating
Body or FEB model, see also Section 5 where ). Considering the
total number of CO molecules in the disk and the CO evaporation
rate of an active comet such as Hale-Bopp, a few times 10$^{11}$
large comets similar to Hale-Bopp would be simultaneously required
to explain the amount of CO gas present in the BP Tau disk. This
is well above the number of FEBs falling on $\beta$~Pic per year
(a few hundred).

Taken together, the unusual mm properties suggest that BP Tau may be a
transient object in the phase of clearing its outer disk.

\subsection{The ambiguous case of HD\,141569}
HD\,141569 is a B9 star located at $\sim $100\,pc. The position of the
star close to the ZAMS in the HR diagram and the presence of an IR
circumstellar excess lead many authors to classify it an HAeBe
star. This was reinforced by the presence of circumstellar gas latter
on detected by Zuckerman \etal (1995) and by the identification of
emission features tentatively attributed to PAH (\cite{sylv96} and
ref. therein) which are frequently observed in SEDs of HAeBe
(e.g.\cite{meeu01}). But, the lack of excess in the NIR, the lack of
photometric variability, the faint intrinsic measured polarimetry
(\cite{yudi00}), and importantly, the low disk to star luminosity
ratio ($8.4\times 10^{-3}$) rather correspond to the description of a
Vega-like star. Both HAeBe and Vega-like classes show a large spread
of ages. With an age of 5\,Myr (Weinberger et al. 2000), HD\,141569
falls at the common edge of the two categories.

HD\,141569 is among the few stars showing a spatially resolved
optically thin dust disk in the NIR. Contrary to the $\beta$ Pictoris
disk, the inclination of the HD\,141569 disk on the line of sight
offers the opportunity to
investigate both the radial and azimuthal profiles of the dust
surface density in great detail. Using coronagraphic techniques,
the HST identified a complex dust structure seen in scattered
light of about ten times our Kuiper Belt size ($\sim 500$~AUs)
(\cite{auge99}). Mid-IR thermal emission observations only partly
compensate the lack of constraints on the innermost regions
($<$1''$\equiv$100\,AU) masked by the HST coronagraph
(\cite{fish00}.) Both data help to sketch out the overall dust
distribution as summarized in Fig.4 from March et al. (2002).  The
dust appears depleted inside $\sim$150\,AUs compared to the outer
regions. The outer disk has a complex shape dominated by two
non-axisymetrical and not accurately concentric wide annulii at
200 and 325\,AU (\cite{moui01}), respectively. Interestingly, the
furthest ring is made of grains smaller than the blow-out size
limit which theoretically points out on the presence of cold gas
in the outer disk(\cite{bocc02}). Surprising, an arc radially thin
but azimuthally extended over $\sim$90$^{\rm o}$, is located at
about 250\,AUs, precisely between the two major ring-like
structures. These informations on the disk morphology are very
valuable because they indicate the impact of internal (planets?)
and/or external gravitational perturbations (stellar companions?,
Augereau \& Papaloizou 2003a).

The detection of a substantial amount of cold gas associated with
an optically thin dust disk is also unusual (\cite{zuck95}).
Recent millimetre interferometric observations of HD\,141569
reveal the gaseous counterpart of the extended disk resolved in
scattered light (Augereau \etal 2004, Fig.\ref{hd141}.)
The CO gas in rotation shows a velocity gradient consistent with
the major axis of the optical disk. Interestingly, hot gas (CO) is
also detected by high resolution mid-IR spectroscopy revealing the
gaseous content of the inner disk (\cite{brit02}), at a few tens
of AUs from the stars.
\begin{figure}
 \caption[]{This montage shows the HST image of HD141569 at 0.5
 $\mu$m from Mouillet et al. (2001) and three channels (redshifted,
 systemic and blueshifted velocity) of CO J=2-1 map from the IRAM
 interferometer. Note that, as expected for rotation the velocity
 gradient of the CO emission is along the major disk axis. From
 Augereau \etal (2004)}
\resizebox{9.0cm}{!}{
}
\label{hd141}
\end{figure}

The HD\,141569 disk possesses NIR properties close to those of the
$\beta$ Pic disk and also Mid-IR and mm properties close to those of
an Herbig Ae disk; as such it seems reasonable to consider it as a
transition disk.

\subsection{The puzzle of Weak Line TTauri Stars}

Weak Line TTauri stars (wTTs) have a SED which presents a weak IR
excess and unlike to CTTs, they do not exhibit strong optical
emission lines (with $W_{H_\alpha} \leq 10$~$\mathrm{\dot{A}}$).
As such, they are usually considered as the evolved counterpart of
the CTTs stars and are classified as class III objects surrounded
by optically thin NIR disk (of a few $\sim 10^7$yr).

However, several studies show that a significant fraction of the
wTTs have ages of same order than those of CTTs (Stahler \& Walter
1993, Grosso \etal 2000). Among them, one interesting example is
the case of V836 Tau: this star presents the optical properties of
wTTs star and in the meantime its mm characteristics are very
similar to those of BP Tau since it is also surrounded by a
compact CO disk (Duvert \etal 2000). Its observed properties are
very similar to those of the BP Tau disk. More recently, Bary
\etal (2002) have reported H$_2$ detection around DOAr 21, a wTTs
located in $\rho$~ Oph which is even more puzzling.

\vskip 0.3cm

We have only a few examples of transition disks and each of them
exhibits very different properties which are also depending on the
observational approach (optical versus mm/submm observations).
This clearly demonstrates that only multi-wavelengths studies can
allow to retrieve the physical properties of these objects. The
scenario by which massive disks dissipate and may form planets is
poorly constrained today.
\section{Planetary Disks}
\label{planeto}
Since the lifetime of massive protoplanetary disks is observed to
be less than a few $\sim 10^7$~years, we {\it a priori} should not
expect disk structure beyond that age. As circumstellar disks
evolve, their mass decreases. When the disks dissipate, the
material become less bright, less dense and apparently more
difficult to detect. IR excess detection of material around nearby
main sequence stars (Auman et al.~1984) leads to the conclusion
that the lifetime of the thin disks is longer than that of massive
disks. The time spent at these late stages being longer, gives the
opportunity to detect evolved disks in the solar neighborhood
(less than $\sim$100~parsecs) around stars older than few
$10^7$~years. These disks are less dense than protoplanetary
disks, but their proximity allows us to observe them in great
detail. The disks seen around main sequence stars are now believed
to be the visible part of more massive systems in which most of
the mass is kept in the form of planetesimals and even planets.
There, planetary formation is either at the end or already
finished (Lagrange \etal 2000).

The duration of the "planetary disk" phenomenon is so long (see
below) that obviously they are by nature not the remaining
material of the protoplanetary disks. It is now clear that these
"planetary disks" has been replenished with material from
pre-existing reservoir. As we will see below, the basic process
needed to sustain these disks is based on the release of dust
and/or gas by colliding asteroids and/or by evaporating
planetesimals. These disks are thus also described as `debris
disks' or `second generation disks'.

\subsection{Dust in Planetary disks}

First detected by IRAS through its infrared excess, the dust
component of planetary disks is the easiest part to detect. The
spectral energy distribution of main sequence stars with
circumstellar material shows infrared excess above $\sim$10$\mu$m
from which it is possible to have some indication on the dust
size, spatial distribution, total mass or simply information on
the fraction of stars harboring such planetary disks (Backman \&
Paresce 1993). More recent ISO surveys of nearby stars show that
about 20\% of the stars have infrared excess attributed to
circumstellar material (Dominik 1999) with typical lifetime of
about 400~million years (Habing et al., 1999, 2001).

\begin{figure}
\resizebox{8.0cm}{!}{\includegraphics[angle=0.0]{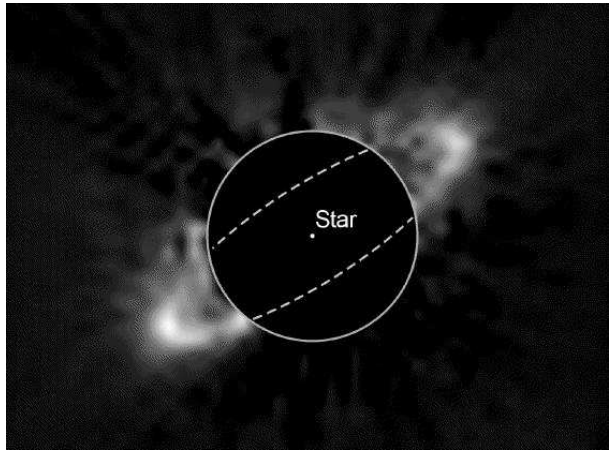}}
\caption{Image of dust scattered light in the HR~4796 ring
observed with the Hubble Space Telescope. Schneider et al.
(1999).} \label{HR4796}
\end{figure}

For an extremely small fraction of these disks, it is possible to
image the dust. These images are produced by the scattered light
at visible wavelengths, or by images of the infrared thermal
emission of the warm part of the disk at 10 or 20$\mu$m. The first
historical image of such disk was the image of the disk of
$\beta$\,Pictoris obtained with coronographic observations of the
scattered light (Smith \& Terrile 1984). Imaging is difficult;
indeed $\beta$\,Pic remained the only disk imaged (see e.g.
Lecavelier des Etangs et al., 1993; Kalas \& Jewitt, 1995;
Mouillet et al., 1997a; Heap et al. 2000) until the late 90's when
new instruments allowed observers to image few other disks by the
detection of the scattered light (Schneider et al., 1999;
Fig.~\ref{HR4796}) or by the detection of the thermal emission in
the infrared (around HR~4796; Jawardhana et al., 1998; Koerner et
al., 1998) or in the sub-millimeter (images of Vega, Fomalhaut and
$\beta$~Pic have been obtained by Holland et al., 1998, and of
$\epsilon$~Eridani by Greaves et al., 1998).

Observations of the dust provide important constraints on the
spatial structure of the disks. For example, the morphology of the
$\beta$~Pictoris disk and the inferred spatial distribution have
been analyzed in great details (Artymowicz et al. 1989; Kalas \&
Jewitt 1995). Images revealed unexpected properties, like the
presence of a break at $\sim$120~AU in the radial distribution of
the dust, the warp of the disk plane, and various asymmetries. All
asymmetric features are often attributed to gravitational
perturbation of massive bodies such as Jupiter-mass planets
(Lecavelier des Etangs et al., 1996; Lecavelier des Etangs, 1997b;
Augereau et al., 2001).

The most intriguing characteristic of the inner part of the
$\beta$\,Pic disk is the so-called `warp', which consists of a
change of the inclination of the mid-plane of the disk inside
about 80~AU. This warp is explained by the presence of a planet on
an inclined orbit with the same inclination as the tilt of the
inner disk (Mouillet et al., 1997b). The measured warp distance
allows to constrain $M_p \times D_p^2 $ where $M_p$ and $D_p$ are
the mass and the distance of the perturbing planet. In the case of
$\beta$\,Pic we have $M_p D_p^2 \approx 2\cdot 10^{-3} M_{*}
(10{\rm AU})^2 (t/10^7 {\rm yr})^{-1}$. If the age of the system
is $t\sim~2\cdot 10^{7}$~years (Barrado y Navascues et al. 1999),
then a Jupiter mass planet at 10~AU and inclined by 5~degrees from
the disk plane can easily explain the observed warp.

Similarly, in the case of HR~4796, sharp truncation of the ring
structure is observed (Fig.~\ref{HR4796}; Schneider et al., 1999). The
presence of a ring is expected to shed light on the physical
phenomena occurring at typical places where planets are supposed
to form. However there is still not a general agreement on the
interpretation of this truncation which could be produced by the
gravitational perturbations of a planet (e.g., Wyatt et al., 1999)
or by drag of the dust by the gas component of the disk (Klahr \&
Lin, 2001; Takeushi \& Artymowicz, 2001).

A key point concerning disks around main sequence stars is that
the dust life-time is shorter than the age of these systems (see
Fig.~8 p.206, Artymowicz, 1997). In the $\beta$\,Pic disk, dust
particles are destroyed by collisions between grains which produce
submicronic debris quickly eliminated by the radiation pressure.
In the less dense disks like the $\epsilon$~Eri ring (Greaves et al.,
1998), the Pointing-Robertson drag is the dominant process
which also eliminates the dust on time-scale shorter than the disk
age. As the dust life-time is very short, one must consider that
the observed dust is continuously resupplied (Backman \& Paresce
1993). It is generally thought that the "debris" disks are
substantial disks of colliding planetesimals (see Backman et al.
1995 for an analogy with collision in the Solar system Kuiper
belt). These disks can thus be considered as the signature of
complete planetary systems which, like our own Solar system,
contain interplanetary dust, asteroid-like kilometer-sized bodies,
and, probably, comets and planets. The visible part of these disks
is only the fraction of material having the largest cross section,
showing the presence of invisible but more massive objects.

\subsection{Gas disks around main sequence stars}

 Although apparently more difficult to interpret, the gas component of the
planetary disks gives the opportunity of the most detailed
modeling of circumstellar processes. In particular, the
$\beta$\,Pic spectroscopic variability is now well explained in
many details by the evaporation of cometary objects close to the
star (see Sect~\ref{betapic}).

In contrary to emission from more massive disks, emission lines
from the gaseous planetary disks are, in most cases, below the
detection limit of current instruments. Molecular transitions at
millimeter wavelengths are too faint to be detected. They give
only upper limits on the gas content (Dent et al. 1995, Liseau
1999). Detections of infrared emission of H$_2$ at 17 and 28$\mu$m
have been claimed with ISO around main sequence stars (Thi et al.,
2001a, 2001b). However these detections have been challenged by
ground based observations at 17$\mu$m which show no detection with
three times better sensitivity (Richter et al., 2002). FUSE
observations also showed that if the ISO detection of H$_2$
emission around $\beta$\,Pic is real, then this H$_2$ is not
distributed widely throughout the disk (Lecavelier des Etangs et
al., 2001). The HST detection of Fe\,{\sc ii} emission lines in
the disk of $\beta$\,Pic is marginal and still to be confirmed
(Lecavelier des Etangs et al. 2000). Finally, the only strong
detection of emission from the gaseous component in a planetary
disk has been performed by Olofsson et al. (2001). With
high-resolution spectroscopy of the $\beta$\,Pic disk, they
clearly detected the resonantly scattered sodium emission through
the Na\,{\sc i} doublet line at 5990 and 5996~\AA. The gas can be
traced from less than 30~AU to at least 140~AU from the central
star. Unfortunately, this observation remains unique. This
definitely opens a new field of observation with an original
technic.

Although emission spectroscopy of tenuous gas disks is difficult,
absorption spectroscopy is much more sensitive. In planetary disks
seen nearly edge-on, the central star can be used as a continuum
source, and the detection of absorption lines offers the
opportunity to scrutinize the gaseous content in details.

\begin{figure}
\resizebox{8.0cm}{!}{\includegraphics[angle=-90.0]{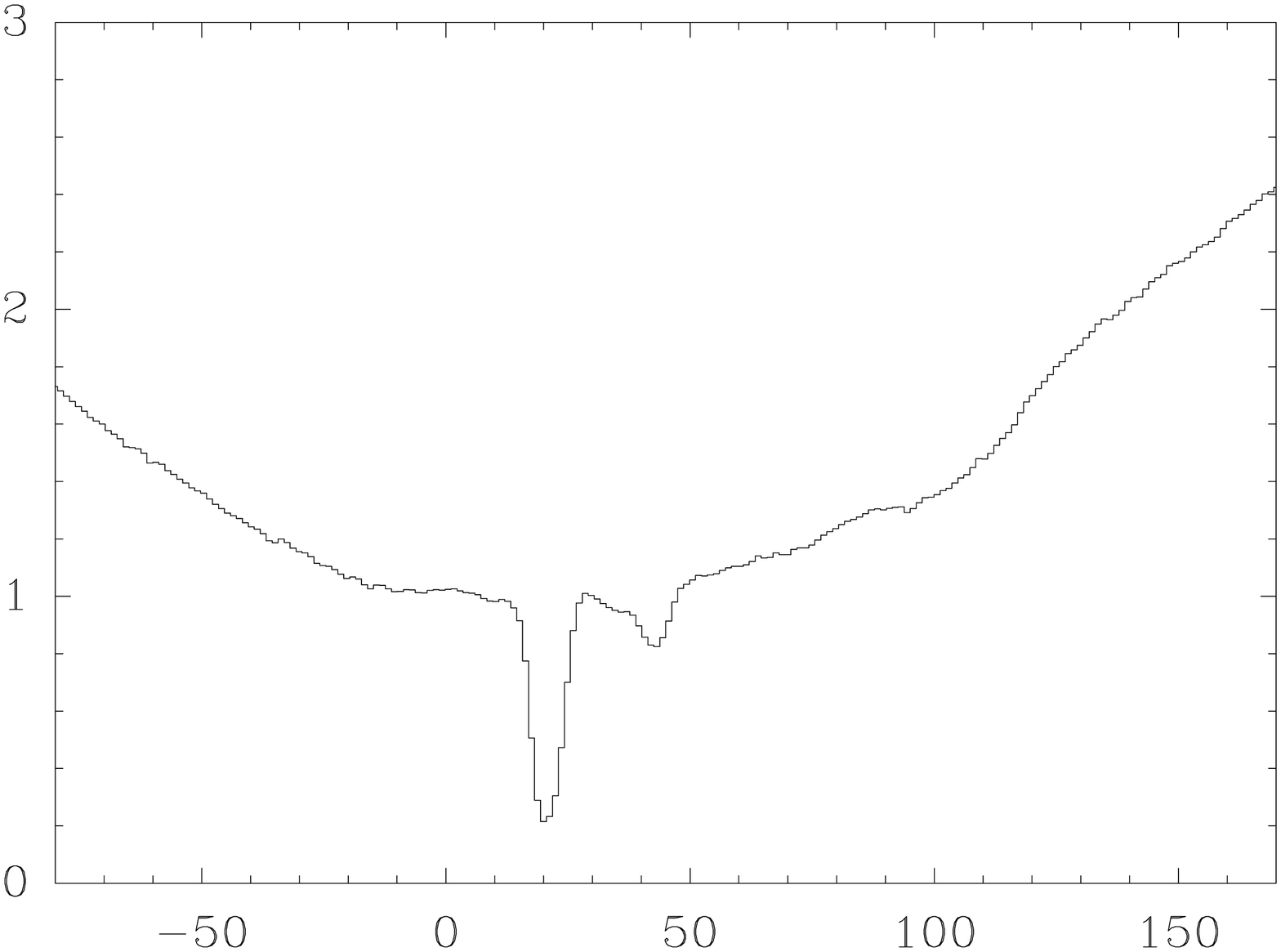}}
\resizebox{8.0cm}{!}{\includegraphics[angle=-90.0]{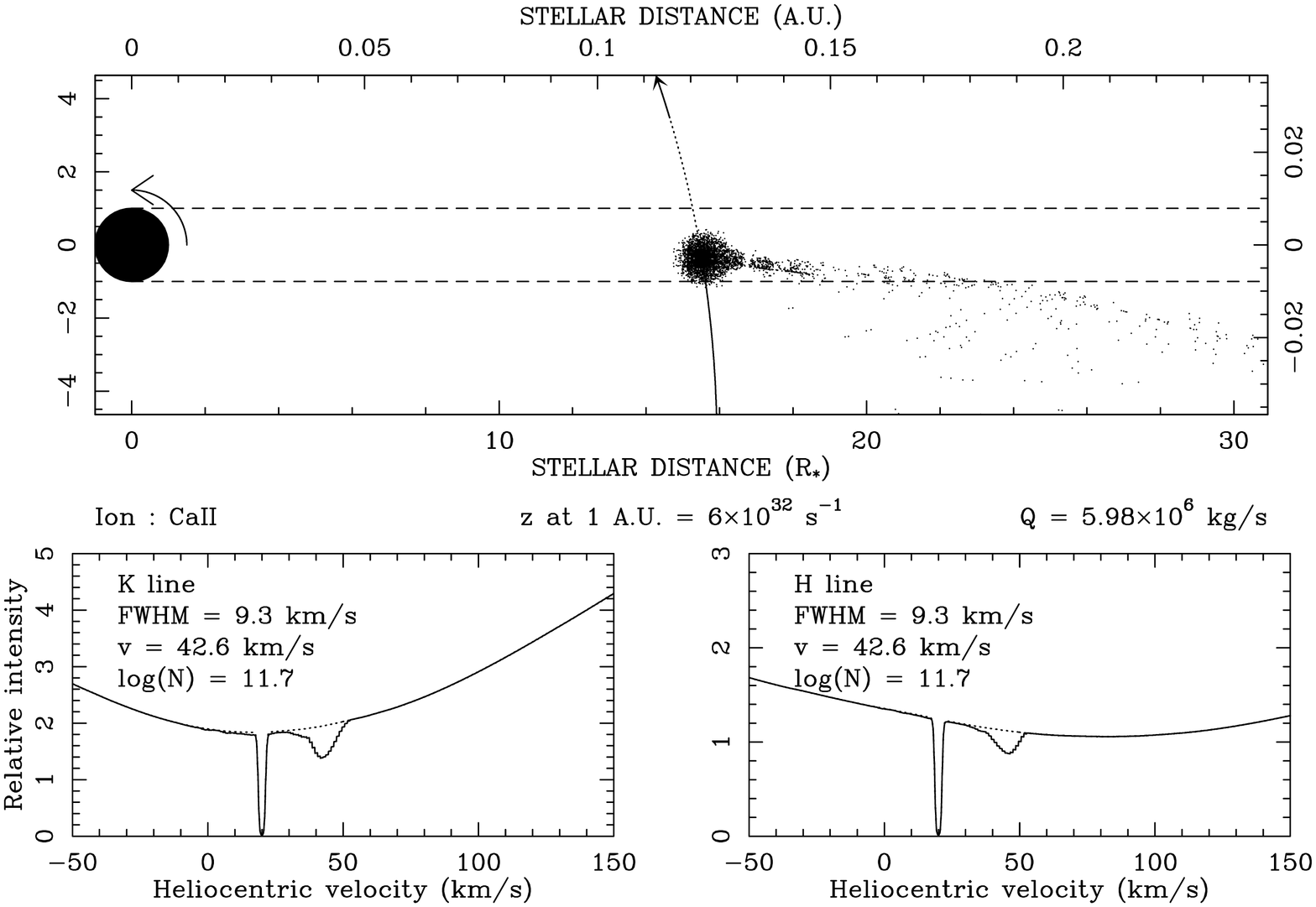}}
\caption[]{ {\bf a}. Ca\,{\sc ii} spectrum of $\beta$\,Pic (Lecavelier
des Etangs 2000). The stable component is visible at 20~km\,s$^{-1}$
heliocentric velocity. This spectrum shows two variable redshifted
components : one sharp absorption at low velocity and one broad
absorption line at high velocity ($>100$~km~s$^{-1}$). {\bf b}.
Simulation of Ca\,{\sc ii}\ absorption lines produced by a FEB at
large distance. The line is redshifted and sharp, with a profile
very similar to the profile of the redshifted variable lines
observed in the $\beta$\,Pic spectrum.} \label{FEB}
\end{figure}

Few months after the discovery of the dust disk around
$\beta$\,Pic, its gaseous counterpart was discovered through the
Ca\,{\sc ii} absorption lines (see Fig.~\ref{FEB}) (Hobbs et al.,
1985; Vidal-Madjar et al., 1986). Because one absorbing component
is seen identically in all observations and at the same radial
velocity as the star (20~km~s$^{-1}$), it is named the ``stable''
gas component. This stable gas is composed of small amounts of
neutral sodium and iron, as well as large amounts of singly
ionized species like Ca\,{\sc ii}, Fe\,{\sc ii}, Mg\,{\sc ii},
Mn\,{\sc ii}, Al\,{\sc ii}. The overall composition is close to
solar (Lagrange et al. 1998). Ultraviolet spectroscopy leads to
the detection of two very peculiar elements : C\,{\sc i} and the
CO molecule, which both have short life-time. In the other hand,
the OH molecule was not detected with a relatively tight upper
limit (Vidal--Madjar et al. 1994). The numerous electronic
transitions of CO in the ultraviolet give constrains on the column
density, temperature ($\sim 20$~K) and a very unusual isotopic
ratio $^{12}$CO/$^{13}$CO=15$\pm$2 (Jolly et al., 1998; Roberge et
al., 2000).

Circumstellar gas signatures similar to the $\beta$\,Pic ones are
also seen around some other main sequence stars. It should be
noticed however that in the rare cases where a Ca\,{\sc ii} (or
e.g., Fe\,{\sc ii}) line at the star radial velocity has been
detected toward other stars like HR~10, these stars have been
identified because they also show either spectral variability, or
the presence of over-ionized species, or redshifted optically
thick absorption lines. This lack of detection of only the
circumstellar absorption at the systemic velocity may be due to
possible confusion with the interstellar medium. The first main
sequence star discovered to have a very similar spectroscopic
behavior as $\beta$\,Pic is HR~10 (Lagrange et al. 1990). This
star shows variable redshifted or blueshifted absorption lines
(Lagrange et al., 1990; Welsh et al., 1998), and a central
component seems relatively stable. Very highly excited levels of
Fe\,{\sc ii} have been detected, proving that the gas is not
interstellar but circumstellar. Redshifted optically thick lines
of Mg\,{\sc ii} have also been detected and interpreted as small
clouds of excited gas falling toward the star (Lecavelier et al.,
1998).

51~Oph is an interesting case because circumstellar dust is
present simultaneously with the gas: 51~Oph presents a complex
system with dusty infrared excess due to cold dust, silicates
emission features, absorption lines by overionized species (Grady
\& Silvis 1993), Fe\,{\sc ii} at excited levels, abnormal Mg\,{\sc
ii} ratio, and finally a possible detection of C\,{\sc i}  in the
circumstellar gas (Lecavelier des Etangs et al., 1997b). Links
between these gas and dust features have still to be understood.

\subsection{The $\beta$\,Pic disk: a cometary disk}
\label{betapic}

The $\beta$\,Pic disk has certainly been the most fruitful for
surprises and discoveries. In addition to the gaseous stable
component, variable absorption features have been detected and
surveyed since 1984 (Fig.~\ref{FEB}). Slowly variable features are
most of the time confined to one or two components redshifted by
10 to 30 km\,s$^{-1}$~relative to the star. Although such
structures seem to be changing in both velocity and strength (by
about $\pm$~10~km\,s$^{-1}$~and large factors in strength), they
nevertheless remain very comparable during few consecutive hours,
often from one day to the next and even sometime over weeks.

Some other components present strong variability, in particular in
the Mg\,{\sc ii}\ and Al\,{\sc iii}\ lines. These components are
also observed in the Ca\,{\sc ii}\ and Fe\,{\sc ii}\ line as weak
and broad absorptions spread over few tens of~km\,s$^{-1}$. The
changes are observed on very short time scale, hours or even less.
These features are mostly strongly redshifted, with shifts that
could reach 300 to 400 km\,s$^{-1}$. These highly varying features
were detected only in ionized species, including highly
(over-)ionized ones like Al\,{\sc iii}\ and C\,{\sc iv}\
completely unexpected in such a relatively ``cool'' stellar
environment. The rapid changes make these features difficult to
track.

These spectral variations are interpreted with a scenario of
star-grazing comets. The presence of strong redshifted ionized gas
is difficult to understand, since the very high radiation pressure
should expel it very quickly. The variable absorption lines are
almost always redshifted, although blueshifted ones would have
been expected. The gas must then be injected with very high inward
radial velocity. There is only one simple way to produce this
situation, namely the evaporation of grains moving toward the
star. Since the radiation pressure acts on grains, they must be
injected with high velocity, through the evaporation from more
massive bodies for which the gravitation is much larger that the
radiation force. This model has been developed in great detail
(Beust~et al. 1990, 1991a, 1991b), and can be summarized by
`evaporation of star-grazing comets'. Indeed, the strong
variability in the circumstellar lines of the ionized elements
like Fe\,{\sc ii}, Al\,{\sc iii}, C\,{\sc iv} and Mg\,{\sc ii} is
now attributed to the evaporation of kilometer size,
``cometary-like'' bodies falling toward the star: this is the
Falling Evaporating Bodies (FEB) scenario. The over-ionized
variable species Al\,{\sc iii} and C\,{\sc iv} cannot be produced
by photoionization, but Beust and Tagger (1993) showed that they
can be formed by collisional ionization in the coma surrounding
these Falling Evaporating Bodies.

Among the different phenomena that can be explained by this model,
one can stress on the observation of abnormal ratio in doublet
lines. For example, although the Mg\,{\sc ii} doublet has an
intrinsic oscillator strength ratio of two, the measured ratio is
exactly one, even for unsaturated lines (Vidal-Madjar et al.
1994). This proves that the absorbing gas cloud is optically thick
(ratio equals to one) but does not cover the total stellar disk
(lines are not saturated). This behavior is directly explained by
the FEB model which produces clumpiness of the absorbing clouds
(Beust~et al. 1989). Similar ratios have also been detected in
redshifted absorptions toward other `$\beta$\,Pic-like stars'.

With several hundreds of FEBs per year, the frequency is several
orders of magnitudes higher than that of sungrazing comets in the
solar system. Planetary perturbations are thought to be the
process responsible. Direct scattering by close encounters with a
massive planet does not seem to be efficient unless the planet
eccentricity is very high (Beust et al. 1991b). Beust \&
Morbidelli (1996) proposed a generic model based on a mean-motion
resonance with a single massive planet on a moderately eccentric
orbit ($e\simeq0.05$). Indeed, a test particle trapped in the 4:1
resonance with such a planet becomes star-grazing after $\simeq
10\,000$ planetary revolutions. This model explains not only the
preferred infall direction but also the radial velocity-distance
relation observed in the FEBs.

Many other stars show redshifted absorption lines (HR~2174, 2~And,
etc.). All these detections of redshifted absorption lines raise
the question of the explanation for the quasi-absence of
blueshifted events. In the $\beta$\,Pic case, it is believed that
the orbits of the star-grazing comets present always about the
same angle to the observer (because the gas is seen in absorption
against the stellar continuum in an edge-on disk). However, when
observed on several stars, some blueshifted orientation should be
expected. Given the very impressive fit between the $\beta$\,Pic
observations and the FEB model, it is likely that the $\beta$\,Pic
FEB phenomenon is somehow particular and that other stars present
either real ``infall'' on the star or their evaporating bodies may
be generally destroyed before they reach the periastron (Grinin et
al. 1996). In the last case, the process needed to put these
bodies on very eccentric orbits in less than one orbital
time-scale remains to be found.

Another class of cometary-like object might also be present around
$\beta$\,Pic. Collision of planetesimals are believed to
continuously resupply most of the dusty disks around main sequence
stars. In the case of $\beta$\,Pic, a significant part of the disk
can be also produced by the evaporation of kilometer-sized bodies
located at several tens of AU from the central star (Lecavelier
des Etangs et al. 1996). Indeed, in the $\beta$\,Pic disk, CO
evaporates below 120~AU from the star. If bodies enter that
region, they start to evaporate and eject dust particles. These
particles are subsequently spread outward in the whole disk by the
radiation pressure. The distribution of their eccentric orbits
gives a dust surface density similar to one observed around
$\beta$\,Pic. This alternative scenario for the production of dust
in the $\beta$\,Pic\ disk easily explains any asymmetry even at
large distances, because a planet in the inner disk can have
influence on the distribution of nearby parent bodies producing
the dust spread outward (Lecavelier des Etangs, 1998).

The observed CO/dust ratio is another argument in favor of this
scenario. An important characteristic of the $\beta$\,Pic\ disk is
the presence of cold CO and C\,{\sc i}\ (Vidal-Madjar et al. 1994,
Roberge et al. 2000). CO and C\,{\sc i}\ are destroyed by
ultraviolet interstellar photons (extreme UV flux from the star is
negligeable). Like the dust, they have lifetime shorter than the
age of the star ($t_{\rm CO}$$\sim$$t_{\rm CI}$$\sim$200~years). A
mechanism must replenish CO with a mass rate of $\dot{M_{CO}} \sim
10^{11} {\rm kg\ s}^{-1}$ The corresponding dust/CO supplying rate
is $\dot{M_{\rm dust}}/\dot{M_{CO}}\approx 1$. This is very
similar to the dust/CO ratio in the material supplied by
evaporation in the solar system. This provides an indication that
the $\beta$\,Pic\ dust disk could be supplied by evaporating
bodies orbiting at several tens of AU from the star like Chiron
evaporates at dozen AU from the Sun.

\subsection{Towards a global picture}

Disks around main sequence stars are probably related to the
presence of young planetary systems in a phase of strong activity.
They show that the planetary systems are still active and evolve
after their formation.

There are still many unknowns. This new field of astronomy is
still a collection of different objects which do not correspond to
an evolutionary scheme. A global picture is still to be built. It
is clear that the extremely numerous names to qualify these disks
which we selected to call the "planetary disks", show that there
is not an unique understanding of the phenomenon. Some authors
refer to "Vega phenomenon", often to qualify the infrared excess.
"$\beta$\,Pic phenomenon" is even more confusing regarding the
number of different phenomena observed around that star.
"Kuiper-disk", "cometary disk", or alternatively "debris disk"
refer to evidence concerning the different origins of these disks.
It is not yet clear if the many pieces shown here correspond to
the same puzzle. New observations and theoretical works will be
needed to solve these issues in the next decade.
\section{Observations in Future}
\label{futur}
The examples given in the previous sections clearly demonstrate
that the frontier between the different classes of objects is not
well constrained, partly because the statistics is still too poor
to provide a quantitative understanding of some of the observed
properties. Since very few disks have been resolved so far it
remains very speculative to derive a timescale and a detailed
evolutionary scheme from protoplanetary to planetary disks.
Moreover, concerning the protoplanetary phase, our understanding
is crudely limited to the cold outer disks ($r \geq 50$~AUs).

Throughout this chapter, the several examples also show that only
multi-wavelength studies of the observational properties will
allow astronomers to properly incorporate in models the physical
processes in action.

\subsection{Challenge for protoplanetary / transition disks}

\begin{figure}
 \caption[]{ This montage summarizes the
observable properties of a protoplanetary disk encountered around
a TTauri star located at 150~pc. This illustrates which region of
the disk is sampled depending on the telescope in
use.}\label{plandisk}
\resizebox{9.0cm}{!}{\includegraphics[angle=270.0]{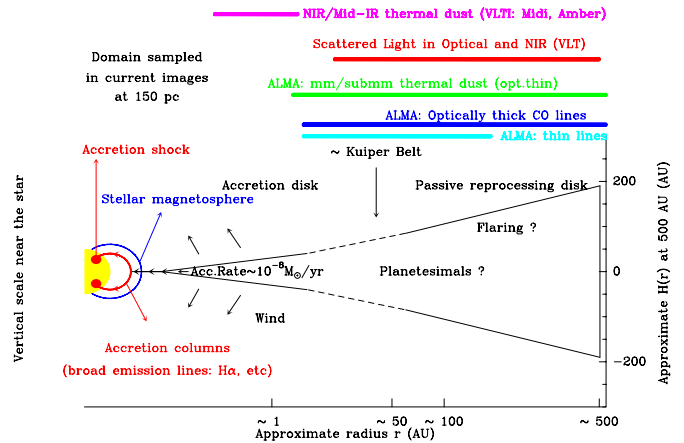}}
\end{figure}

In protoplanetary disks, most of the material lies in the
relatively cold outer part of the disks. Hence, resolved sub-mm
observations, obtained with large mm arrays such as ALMA, will
provide the best tool to investigate this reservoir of mass ($r>
20-50$~AUs). In particular, ALMA will observe large samples,
allowing statistics on disk properties and frequency. Of course,
to study the hotter inner disk, where planets form ($0.1 \geq r
\geq 10$~AUs), optical, NIR and Mid-IR interferometry techniques
are required. As soon as they will be able to produce images (even
with a few baseline numbers), instruments such as AMBER and MIDI
on the VLTI or OHANA will add to our understanding of the dust
properties and composition. Images (or a reasonable $uv$ coverage)
are required to decide between all the existing models of dust
disks, in particular they should allow us to disentangle between
geometry, temperature and opacity effects. Fig.\ref{plandisk}
summarizes which part of disk can be investigated depending on the
instrument in use. A combination similar of ``ALMA \& VLTI'' would
be very efficient to sample the global disk properties.
A necessary step to understand how planets form is to view gaps
created by protoplanets. For this purpose images are definitely
required either at sub-mm or NIR wavelengths. In its large
configuration, ALMA will have baselines up to 14 km, providing an
angular resolution of $\sim$ 0.03$''$ (or 4~AUs at the Taurus
distance) at $\lambda = 1.3 \mu$m. Hydrodynamics coupled to
radiative transfer simulations of the dust emission at 350~GHz by
Wolf \etal (2002) show that ALMA will be able to resolve out a gap
created by a proto-Jupiter and located at 5~AU from a star at
D=150~pc.
Concerning the gas content, in spectral lines near
$\lambda=1.3$mm, ALMA will be about 30 times more sensitive than
the IRAM array and quantitative chemical studies could begin.
Multi-transitions analysis would even allow observers to measure
abundance gradients in the disk. Protoplanetary disks are indeed
H$_2$ disks and direct investigation of the H$_2$ distribution and
mass remains the more direct way to study how protoplanetary disks
dissipate. This domain will strongly benefit from satellites such
as SIRTF and JWST.
Finally, the knowledge of protoplanetary disks is today biased by
sensitivity and we image only the brighter disks. Disk clearing is
poorly constrained. The ALMA sensitivity will allow to image in the
sub-mm domain many other objects similar to BP Tau and even optically
thin dust disks in the NIR.
\subsection{Challenge for planetary disks}
By extrapolation from ISO results on the occurence of debris disks
around MS stars and according to the Hipparcos catalogue, one can
predict $\sim 10^2$ and $\sim 10^3$ planetary dust disks around
(hypothetical) stars younger than about 0.5\,Gyr within 20\,pc and
50\,pc radii of the Sun respectively. These stars are close enough to
be not limited by angular resolution considerations but their disks
are simply too tenuous to have a chance to be detected with current
instruments. This points up a crucial need for an enhancement in
sensitivity combined with high angular resolution techniques. Precise
disks shapes, fine structures and asymmetries as revealed by imaging
may indeed be signposts of undetected gravitational companions such as
planets perturbing the underlying disk of km-sized bodies that release
the observed short-lived dust.

The observational techniques discussed below are summarized in
Fig.\ref{planpla}. In the NIR, high contrast imaging with single
aperture telescopes is required to detect faint dust disks very
close to a bright star. For instance new generations of adaptive
optics systems and new concepts of coronagraphic masks are under
study with the prime goal of being able to detect faint objects
from the ground, ideally down to planets (e.g. VLT/Planet Finder).
At these wavelengths but also in the mid-IR, the innermost regions
of planetary disks will nevertheless remain unreachable without
the help of interferometry (e.g. KeckI and VLTI). Resolving the
material within the very first AU around young MS stars and
ultimately producing images by NIR and mid-IR interferometry are
attractive challenges in the near future. In the mid-IR, single
aperture telescopes suffer an unavoidable decrease of the spatial
resolution. The predominant gain at these wavelengths will mostly
come from future spaced-based telescopes, especially the 6-\,m
JWST and SIRTF with increases of two or three orders of magnitudes
in detection thresholds. While current high resolution imagers in
the mid-IR are limited to disks around close-by stars younger than
a few 10\,Myr, JWST/MIRI should resolved debris disks around Gyr
old A-type stars at 50\,pc. In the mm, ALMA will permit the
detection of an unresolved (but optically thin) clump of dust of
$10^{-2}$M$_{\oplus}$ orbiting a star up to 100\,pc from us in one
hour of observing time (see also ``ALMA science case'').

\begin{figure}
\caption[]{Same as Fig.\ref{plandisk} but for a planetary disk around
a young MS star such as $\beta$ Pictoris or HR\,4796. A distance of
50\,pc is assumed. The Gaussians give an estimate of the angular
resolution.}\label{planpla}
\resizebox{9.0cm}{!}{
}
\end{figure}

Gas will also certainly concentrate more and more observational
efforts in order to provide a better understanding of its
timescales and dissipation processes. Since the gas content of
young planetary disks is yet badly constrained, anticipating
future results might be very speculative. However, the ALMA
sensitivity would allow to detect, in 1 hour of observing time, a
CO column density of a few $10^{12}$cm$^{-2}$ in a beam of 2.5$''$
(planetary disks are close to us hence angularly extended)
assuming a linewidth of 3 km/s. This is well below the CO column
density of 10$^{15}$cm$^{-2}$ detected by Vidal-Madjar \etal
(1994) in $\beta$ Pic. Depending on the gas geometry, the distance
and the evolutionary status of the disk, ALMA would allow
astronomers to put some important constrains on various gas model
distributions in planetary disks.
\begin{acknowledgements}
We thank A.Vidal-Madjar for fruitful comments. Fig.~\ref{FEB} has
been reproduced with the kind permission of H.~Beust. A.~Dutrey
would like to thank S.~Guilloteau for a long term and fruitful
collaboration. M.~Simon is also acknowledged for a careful reading
of the manuscript. Many thanks to J.~Pety who provided material
for HH30 (Fig.\ref{hh30} and recent mm results).
J.C.Augereau thanks the CNES for financial support. \\

\end{acknowledgements}

\begin{thebibliography}{}

\bibitem{adams}
Adams F.C., Lada C.J., Shu F.H., 1987, ApJ, 312, 788

\bibitem{alma}
ALMA science case: http://www.eso.org/projects/alma/science/

\bibitem{dAle98}
D'Alessio P., et al., 1998, ApJ, 500, 411

\bibitem{dAle99}
D'Alessio P., et al., 1999, ApJ 527, 893.\

\bibitem{Arty89}
Artymowicz P., Burrows C., and Paresce F., 1989,
ApJ 337, 494-513.\

\bibitem{Arty97}
Artymowicz P.\, 1997
{\it Annual Review of Earth and Planetary Sciences, 25, }175

\bibitem[Augereau et al., 1999]{auge99} Augereau J.~C., Lagrange
A.~M., Mouillet D., \& M{\' e}nard F.\, 1999, A\&A, 350, L51

\bibitem{auge01a}
Augereau J.~C., Lagrange A.~M., Mouillet D., \& M{\' e}nard F.\,
2001a, A\&A, 365, 78

\bibitem{Aug01b}
Augereau J.~C., Nelson R.~P., Lagrange A.~M., Papaloizou J.~C.~B.,
and Mouillet D.\, 2001b
A\&A, 370, 447-455.\

\bibitem{auge03a}
Augereau J.~C., \& Papaloizou J., 2003a, A\&A, 414, 1153

\bibitem[Augereau et al., 2004]{auge02} Augereau J.~C., Dutrey A.,
Lagrange A.~M., Mouillet D., \& Forveille T.\, 2004, in
preparation

\bibitem{Aumann}
Aumann H.~H.\, 1985
{\it Publications of the Astronomical Society of the
Pacific, 97, }885-891.\

\bibitem{Back93}
Backman D.E. and Paresce F., 1993
{\it Protostars and Planets III,} 1253--1304.\

\bibitem{Back95}
 Backman D.~E., Dasgupta A., and Stencel R.~E.\, 1995
ApJ, 450, L35

\bibitem{Barr}
Barrado y Navascu{\' e}s D., Stauffer J.~R., Song I., and
Caillault J.-P.\, 1999
ApJ, 520, L123-L126.\

\bibitem{bary}
Bary J.S., Weintraub D.A., Kastner J.H., 2002, ApJ, 576, L73

\bibitem{beck90}
Beckwith, S.V.W., et al., 1990, AJ, 99, 924

\bibitem[Beckwith et al., 2000]{bec00} Beckwith S.V.W., Henning T.,
NakagawaY., in {\it Protostars and Planets IV}, Tucson, University
of Arizona Press, eds. Mannings V., Boss A.P., Russel S.S., p 533

\bibitem{Beust90}
Beust H., Vidal-Madjar A., Ferlet R., and Lagrange-Henri A.~M.\,
1990
A\&A, 236, 202-216.\

\bibitem{Beust91a}
Beust H., Vidal-Madjar A., Ferlet R., and Lagrange-Henri A.~M.\,
1991a
A\&A, 241, 488-492.\

\bibitem{Beust91b}
Beust H., Vidal-Madjar A., and Ferlet R.\, 1991b
A\&A, 247, 505-515.\

\bibitem{Beust93}
Beust H., and Tagger M.\, 1993
{\it Icarus, 106, }42

\bibitem{Beust96}
Beust H., and Morbidelli A.\, 1996
{\it Icarus,120, }358-370.\

\bibitem{Beust98}
Beust H., Lagrange A.-M., Crawford I.~A., Goudard C., Spyromilio
J., and Vidal-Madjar A.\, 1998
A\&A, 338, 1015-1030.\

\bibitem[Boccaletti et al., 2003]{bocc02} Boccaletti A., Augereau
J.~C., Marchis F., \& Hahn J.\, 2003, ApJ, 585, 494

\bibitem{bou}
Bouwman J., de Koter A., van den Ancker M. E., Waters L. B. F. M.,
A\&A, 360, 213

\bibitem[Brittain \& Rettig 2002]{brit02} Brittain S.~D.~\&
Rettig T.~W.\, 2002, Nature, 418, 57

\bibitem{bur}
Burrows C.J. et al., 1996, ApJ, 473, 437

\bibitem{chia97}
Chiang E.I. \& Goldreich, P., 1997, ApJ 490, 368.

\bibitem{close}
Close L., et al., 1998, ApJ, 499, 883

\bibitem{dar02}
Dartois E., et al., 2002, A\&A, 394, 1057

\bibitem{dar03}
Dartois E., Dutrey A., Guilloteau S., 2003, A\&A 399, 778


\bibitem{Dent}
Dent W.~R.~F., Greaves J.~S., Mannings V., Coulson I.~M., and
Walther D.~M.\, 1995
{\it Monthly Notices of the Royal Astronomical Society, 277,
}L25-L29.\

\bibitem{Dom99}
Dominik C.\, 1999
{\it Solid Interstellar Matter: The ISO Revolution},
d'Hendecourt~L., Joblin~C., and Jones~A. Eds, EDP Sciences and
Springer-Verlag, p.277

\bibitem{dull}
Dullemond C.P., Dominik, C., Natta A., 2001, ApJ, 560, 957

\bibitem{dutr94}
Dutrey A., Guilloteau S. \& Simon M., 1994, A\&A 286, 149

\bibitem{dutr96}
Dutrey A., et al., 1996, A\&A 309, 493 (D96)

\bibitem{dutr97}
Dutrey A., Guilloteau S. \& Gu\'elin M., 1997, A\&A 317, L55

\bibitem{dutr03a}
Dutrey A., Guilloteau S., Simon M., 2003, A\&A, 402, 1003.


\bibitem{Duvert}
Duvert G., et al., 2000, A\&A, 355, 165

\bibitem{ed94}
Edwards S., et al., 1994, AJ, 108, 1056

\bibitem[Fisher et al., 2000]{fish00} Fisher R.~S., Telesco
C.~M., Pi{\~ n}a R.~K., Knacke R.~F., \& Wyatt M.~C.\, 2000,
ApJLet., 532, L141

\bibitem{Freud}
Freudling W., Lagrange A.-M., Vidal-Madjar A., Ferlet R., and
Forveille T.\, 1995
\ A\&A, 301, 231

\bibitem{Glas}
Glassgold A.E., et al., 1997, ApJ, 485, 920

\bibitem{Grady}
Grady C.A., and Silvis J.~M.~S.\, 1993, ApJ, 402, L61-L64.\

\bibitem{grad}
Grady C.A., et al., 1999, ApJ, 523, L151

\bibitem{grad03}
Grady C.A., et al., 2003, BASS, 34, 1137

\bibitem{Greaves}
Greaves J.~S., Holland W.~S., Moriarty-Schieven G., Jenness T.,
Dent W.~R.~F., Zuckerman B., McCarthy C., Webb R.~A., Butner
H.~M., Gear W.~K., and Walker H.~J.\, 1998, ApJ, 506, L133-L137.\

\bibitem{Grin96}
Grinin V., Natta A., and Tambovtseva L.\, 1996, A\&A, 313,
857-865.\

\bibitem{grosso}
Grosso N., Montmerle T., Bontemps S., André P., Feigelson E. D.,
A\&A, 2000, 359, 113

\bibitem{guil98}
Guilloteau S. \& Dutrey A., 1998, A\&A 339, 467

\bibitem{gull}
Gullbring E., et al., 1998, ApJ, 492, 323

\bibitem{Hab99}
Habing H.~J., Dominik C., Jourdain de Muizon M., Kessler M.~F.,
Laureijs R.~J., Leech K., Metcalfe L., Salama A., Siebenmorgen R.,
and Trams N.\, 1999
{\it Nature, 401,
}456-458.\

\bibitem{Hab01}
Habing H.~J., Dominik C., Jourdain de Muizon M., Laureijs R.~J.,
Kessler M.~F., Leech K., Metcalfe L., Salama A., Siebenmorgen R.,
Trams N., and Bouchet P.\, 2001
A\&A, 365, 545-561.\

\bibitem{harp}
Harper D.A., et al., 1984, ApJ, 285, 808


\bibitem{hart96}
Hartmann L. \& Kenyon S.J., 1996, ARAA, 34, 207

\bibitem{harta98}
Hartmann L., Calvet N., Gullbring E., D'Alessio P., 1998, ApJ,
495, 385

\bibitem{hartb98}
Hartmann L., 1998, `` Accretion Processes in Star Formation'',
Cambridge University Press (Cambridge Astrophysics series; 32)

\bibitem{Heap}
Heap S.~R., Lindler D.~J., Lanz T.~M., Cornett R.~H., Hubeny I.,
Maran S.~P., and Woodgate B.\, 2000
ApJ, 539, 435-444.\
\bibitem{Henn}
Henning T., Stognienko R., 1996, A\&A, 311, 291


\bibitem{hob85}
Hobbs L.~M., Vidal-Madjar A., Ferlet R., Albert C.~E., and Gry
C.\, 1985
ApJ, 293, L29-L33.\

\bibitem{Hol98}
Holland W.~S., Greaves J.~S., Zuckerman B., Webb R.~A., McCarthy
C., Coulson I.~M., Walther D.~M., Dent W.~R.~F., Gear W.~K., and
Robson I.\, 1998
{\it Nature, 392, }788-790.\

\bibitem{Jaya98}
Jayawardhana R., Fisher S., Hartmann L., Telesco C., Pina R., and
Fazio G.\, 1998 A Dust Disk Surrounding the Young A Star HR
4796A.\ ApJ, 503, L79

\bibitem{john}
Johnstone D., Hollenbach D., Bally J., 1998, ApJ, 499, 758

\bibitem{Jol98}
Jolly A., McPhate J.~B., Lecavelier A., Lagrange A.~M., Lemaire
J.~L., Feldman P.~D., Vidal Madjar A., Ferlet R., Malmasson D.,
and Rostas F.\, 1998
A\&A, 329, 1028-1034.\

\bibitem{kal95}
Kalas P., and Jewitt D.\, 1995
AJ, 110, 794

\bibitem{kas97}
Kastner J., et al., 1997, {\it Science}, 277, 61

\bibitem{key}
Kenyon S. J. \& Hartmann L., 1987, ApJ 323, 714.

\bibitem{kla01}
Klahr H.~H., and Lin D.~N.~C.\, 2001
ApJ, 554, 1095-1109.\

\bibitem{koer93}
Koerner D.W., Sargent A.I.\& Beckwith S.W.W., 1993, {\it Icarus,
106}, 2\

\bibitem{koer97}
Koerner D.W., 1997, {\it Origins of Life and Evolution of the
Biosphere}, 27, 157

\bibitem{lag98}
Koerner D.~W., Ressler M.~E., Werner M.~W., and Backman D.~E.\,
1998
ApJ, 503, L83

\bibitem{krueg}
Kr\"ugel E. \& Siebenmorgen R.\, 1994, A\&A 288, 929.

\bibitem{lag90}
Lagrange-Henri A.~M., Beust H., Ferlet R., Vidal-Madjar A., and
Hobbs L.~M.\, 1990
A\&A, 227, L13-L16.\

\bibitem{lag98}
Lagrange A.-M., Beust H., Mouillet D., Deleuil M., Feldman P.~D.,
Ferlet R., Hobbs L., Lecavelier Des Etangs A., Lissauer J.~J.,
McGrath M.~A., McPhate J.~B., Spyromilio J., Tobin W., and
Vidal-Madjar A.\, 1998
A\&A, 330, 1091-1108.\

\bibitem[Lagrange et al., 2000]{lag00} Lagrange A.M., Backman D.E.,
Artymowicz P., in {\it Protostars and Planets IV}, Tucson,
University of Arizona Press, eds. Mannings V., Boss A.P., Russel
S.S., p 639

\bibitem{lec93}
Lecavelier Des Etangs A., Perrin G., Ferlet R., Vidal Madjar A.,
Colas F., Buil C., Sevre F., Arlot J.~E., Beust H., Lagrange Henri
A.~M., Lecacheux J., Deleuil M., and Gry C.\, 1993
A\&A, 274, 877

\bibitem{lec96}
Lecavelier Des Etangs A., Vidal-Madjar A., and Ferlet R.\, 1996
A\&A, 307, 542-550.\

\bibitem{lec97a}
Lecavelier Des Etangs A., Vidal-Madjar A., Backman D.~E., Deleuil
M., Lagrange A.-M., Lissauer J.~J., Ferlet R., Beust H., and
Mouillet D.\, 1997a
A\&A, 321, L39-L42.\

\bibitem{lec97b}
Lecavelier Des Etangs A., Deleuil M., Vidal-Madjar A.,
Lagrange-Henri A.-M., Backman D., Lissauer J.~J., Ferlet R., Beust
H., and Mouillet D.\, 1997b
A\&A, 325, 228-236.\

\bibitem{lec98}
Lecavelier Des Etangs A.\, 1998
A\&A, 337, 501-511.\

\bibitem{leca}
Lecavelier des Etangs A., 2000, in {\it Disk, Planetesimals, and
Planets}, ASP Conf. Ser. 219, page 308.

\bibitem{lec00}
Lecavelier des Etangs A., Hobbs L.~M., Vidal-Madjar A., Beust H.,
Feldman P.~D., Ferlet R., Lagrange A.-M., Moos W., and McGrath
M.\, 2000
A\&A, 356, 691-694.\

\bibitem{lec01}
Lecavelier des Etangs A., Vidal-Madjar A., Roberge A., Feldman
P.~D., Deleuil M., Andr{\' e} M., Blair W.~P., Bouret J.-C., D{\'
e}sert J.-M., Ferlet R., Friedman S., H{\' e}brard G., Lemoine M.,
and Moos H.~W.\, 2001
{\it Nature, 412, }706-708.\

\bibitem{lec}
Lecavelier des Etangs A., Deleuil M., Vidal-Madjar A., Roberge A.,
Le Petit F., H\'ebrard G., Ferlet R., Feldman P.D., D\'esert
J.-M., Bouret, J.-C. 2003, A\&A, 407, 935

\bibitem{lei}
Leinert Ch., Haas M., Abraham P., Richichi A., 2001, A\&A, 375,
927

\bibitem{lis99}
Liseau R.\, 1999
A\&A, 348, 133-138.\


\bibitem{Mann97}
Mannings V., \& Sargent A.I., 1997, ApJ 490, 792.

\bibitem[Marsh et al., 2002]{mars02} Marsh K.~A.,
Silverstone M.~D., Becklin E.~E., Koerner D.~W., Werner M.~W.,
Weinberger A.~J., \& Ressler M.~E.\, 2002, ApJ, 573, 425

\bibitem[Mathieu et al., 2000]{mat00} Mathieu R. D., Ghez A. M.,
Jensen E. L. N., Simon M., in {\it Protostars and Planets IV},
Tucson; University of Arizona Press, eds. Mannings V., Boss A.P.,
Russel S.S., p 703


\bibitem{McCabe}
McCabe C., Duchêne G., Ghez A. M., 2002, ApJ, 575, 974

\bibitem{mccaug}
McCaughrean M.J. \& O'Dell C.R., 1996, AJ, 111, 1977

\bibitem[Meeus et al., 2001]{meeu01} Meeus G., Waters
L.~B.~F.~M., Bouwman J., van den Ancker M.~E., Waelkens C., \&
Malfait K.\, 2001, A\&A, 365, 476

\bibitem{Mon}
Monnier J., Millan-Gabet R., 2002, ApJ 579, 694

\bibitem{mou97a}
Mouillet D., Lagrange A.-M., Beuzit J.-L., and Renaud N.\, 1997a
A\&A, 324, 1083-1090.\

\bibitem{mou97b}
Mouillet D., Larwood J.~D., Papaloizou J.~C.~B., and Lagrange
A.~M.\, 1997b
{\it Monthly Notices of the Royal Astronomical Society, 292, }896

\bibitem[Mouillet et al., 2001]{moui01}
Mouillet D., Lagrange A.~M., Augereau J.~C., \& M{\' e}nard F.\,
2001, A\&A, 372, L61

\bibitem{naj}
Najita J.R., Bergin E.A., Ullom, J. N., 2001, ApJ, 561, 880

\bibitem[Najita et al., 2000]{naj00} Najita J.R., Edwards S., Basri G., Carr J.,
in {\it Protostars and Planets IV}, Tucson; University of Arizona
Press, eds. Mannings V., Boss A.P., Russel S.S., p 457

\bibitem{natta}
Natta A., Grinin V., Mannings V., 2000, in {\it Protostars and
Planets IV}, p559

\bibitem{nattab}
Natta A., \& Whitney B. A., 2000, A\&A, 364, 633

\bibitem{olof01}
Olofsson G., Liseau R., and Brandeker A.\, 2001
ApJ, 563, L77-L80.\

\bibitem{pety}
Pety J., et al., 2003, in prep.

\bibitem{pie}
Pietu V., Dutrey A., Khahane C., 2003, A\&A, 398, 565

\bibitem{poll}
Pollack J., et al., 1994, ApJ 421, 615

\bibitem{prin81}
Pringle J.E. 1981, AARA 19,137

\bibitem{rich02}
Richter M.~J., Jaffe D.~T., Blake G.~A., and Lacy J.~H.\, 2002
ApJ, 572, L161-L164.\

\bibitem{rob00}
Roberge A., Feldman P.~D., Lagrange A.~M., Vidal-Madjar A., Ferlet
R., Jolly A., Lemaire J.~L., and Rostas F.\, 2000
ApJ, 538, 904-910.\

\bibitem{rod96}
Roddier C., et al., 1996, ApJ, 463, 326

\bibitem[Rucinski(1985)]
Rucinski, S.~M.\ 1985, AJ, 90, 2321

\bibitem{schnei99}
Schneider G., Smith B.~A., Becklin E.~E., Koerner D.~W., Meier R.,
Hines D.~C., Lowrance P.~J., Terrile R.~J., Thompson R.~I., and
Rieke M.\, 1999
ApJ, 513, L127-L130.\

\bibitem{shu}
Shu, F.H., Adams, F.C., Lizano, S., 1987, ARA\&A 25, 23

\bibitem{shak73}
Shakura N.I, \& Sunyaev R.A., 1973, A\&A, 24, 337

\bibitem{simo00}
Simon M., Dutrey A. \& Guilloteau S.\, 2000, ApJ 545, 1034.

\bibitem{smter}
Smith B.~A., and Terrile R.~J.\, 1984
{\it Science, 226, }1421-1424.\

\bibitem{Sta}
Stahler S.W., \& Walter F.M., 1993, in {\it Protostars \& Planets
III}, 405

\bibitem{Stap95}
Stapelfeldt K.R. et al., 1995, BAAS, 187, 113.04

\bibitem[Sylvester \& Skinner 1996 ]{sylv96} Sylvester
R.~J.~\& Skinner C.~J.\, 1996, MNRAS, 283, 457

\bibitem{take01}
Takeuchi T., and Artymowicz P.\, 2001
ApJ, 557, 990-1006.\

\bibitem{thi01a}
Thi W.~F., Blake G.~A., van Dishoeck E.~F., van Zadelhoff G.~J.,
Horn J.~M.~M., Becklin E.~E., Mannings V., Sargent A.~I., van den
Ancker M.~E., and Natta A.\, 2001
{\it Nature, 409, }60-63.\

\bibitem{thi01b}
Thi W.~F., van Dishoeck E.~F., Blake G.~A., van Zadelhoff G.~J.,
Horn J., Becklin E.~E., Mannings V., Sargent A.~I., van den Ancker
M.~E., Natta A., and Kessler J.\, 2001
ApJ, 561, 1074-1094.\

\bibitem{thi02}
Thi W.F., et al., 2002, A\&A, 394, L27

\bibitem[van den Ancker et al., 1998]{vand98} van den Ancker M.~E.,
de Winter D., \& Tjin A Djie H.~R.~E.\, 1998, A\&A, 330, 145

\bibitem{zad01}
van Zadelhoff G.-J., et al., A\&A, 377, 566

\bibitem{vidal86}
Vidal-Madjar A., Ferlet R., Hobbs L.~M., Gry C., and Albert C.~E.\
1986
A\&A, 167, 325-332.\

\bibitem{vidal94}
Vidal-Madjar A., Lagrange-Henri A.-M., Feldman P.~D., Beust H.,
Lissauer J.~J., Deleuil M., Ferlet R., Gry C., Hobbs L.~M.,
McGrath M.~A., McPhate J.~B., and Moos H.~W.\, 1994
A\&A, 290, 245-258.\

Protostars and Planets IV (Book - Tucson: University of Arizona
Press; eds Mannings, V., Boss, A.P., Russell, S. S.), p. 639

\bibitem{Weid}
Weidenschilling 1997,

\bibitem[Weinberger et al., 1999]{wein99} Weinberger A.~J.,
Becklin E.~E., Schneider G., Smith B.~A., Lowrance P.~J.,
Silverstone M.~D., Zuckerman B., \& Terrile R.~J.\, 1999, ApJLet.,
525, L53

\bibitem{wael}
Waelkens C., Malfait K., Waters L. B. F. M., 1999, EM\&P, 79, 265

\bibitem[Weinberger et al., 2000]{wein00} Weinberger A.~J.,
Rich R.~M., Becklin E.~E., Zuckerman B., \& Matthews K.\, 2000,
ApJ, 544, 937

\bibitem{welsh}
Welsh B.~Y., Craig N., Crawford I.~A., and Price R.~J.\, 1998
A\&A, 338, 674-682.\

\bibitem{wolf}
Wolf S., Gueth F., Henning T., Kley W., 2002, ApJ, 566, L97

\bibitem{wri}
Wright E., 1987, AJ, 320, 818

\bibitem{wya}
Wyatt M.~C., Dermott S.~F., Telesco C.~M., Fisher R.~S., Grogan
K., Holmes E.~K., and Pi{\~ n}a R.~K.\, 1999
ApJ, 527, 918-944.\


\bibitem[Yudin 2000]{yudi00} Yudin R.~V.\, 2000, A\&ASupp., 144, 285

\bibitem[Zuckerman et al., 1995]{zuck95}
Zuckerman B., Forveille T., \& Kastner J.~H.\, 1995, \it {Nature,
373}, 494

\end{thebibliography}
\end{document}